\documentclass[journal=jacsat,manuscript=article]{achemso}

\usepackage[version=3]{mhchem} 
\usepackage[T1]{fontenc}       
\usepackage[utf8]{inputenc}
\usepackage{makecell}
\usepackage{multirow}
\usepackage{tabu}

\author{Heigo Ers}
\affiliation[UT]{Institute of Chemistry, University of Tartu, Ravila 14a, Tartu 50411, Estonia}
\author{Meeri Lembinen}
\affiliation[UT]{Institute of Physics, University of Tartu, Ostwaldi 1, Tartu, 50411, Estonia}
\author{Maksim Mi\v{s}in}
\affiliation[UT]{Institute of Chemistry, University of Tartu, Ravila 14a, Tartu 50411, Estonia}
\author{Ari~P.~Seitsonen}
\affiliation[ENS]
{D\'{e}partement de Chimie, \'{E}cole Normale Sup\'{e}rieure, 24 rue Lhomond, F-75005 Paris, France}
\alsoaffiliation[CSP]
{Centre National de la Recherche Scientifique, Paris Sciences et Lettres, Sorbonne Universit\'{e}, France}
\author{Maxim~V.~Fedorov}
\affiliation[Skolkovo]{Skolkovo Institute of Science and Technology, Moscow, Russia}
\alsoaffiliation[Strath]{Department of Physics, Scottish Universities Physics Alliance (SUPA), Strathclyde University, John Anderson Building, 107 Rottenrow East, G4 0NG, Glasgow, UK}
\author{Vladislav B. Ivani\v{s}t\v{s}ev}
\email{vladislav.ivanistsev@ut.ee}
\affiliation[UT]{Institute of Chemistry, University of Tartu, Ravila 14a, Tartu 50411, Estonia}

\title
  {Graphene--ionic liquid interfacial potential drop from DFT-MD simulations}

\begin{document}



\section*{Abstract}

Ionic liquids (IL) are promising electrolytes for electrochemical applications due to their remarkable stability and high charge density.
Molecular dynamics simulations are essential for better understanding the complex phenomena occurring at the electrode--IL interface. 
In this work, we have studied the interface between graphene and 1-ethyl-3-methyl-imidazolium tetrafluoroborate IL, using density functional theory-based molecular dynamics simulations at variable surface charge densities. 
We have disassembled the electrical double layer potential drop into two main components: one involving atomic \textit{charges} and the other -- \textit{dipoles}.
The latter component arises due to the electronic polarisation of the surface and is related to concepts hotly debated in the literature, such as the Thomas--Fermi screening length, effective surface charge plane, and quantum capacitance.


\section{Introduction}

The ever-growing, world energy consumption has created a demand for higher capacity energy storage devices.\cite{macfarlane_energy_2014} 
Supercapacitors were developed to meet that demand by storing energy in the electrical double layer (EDL) -- at the interface between an electrode and an electrolyte.\cite{conway_electrochemical_1999,kotz_principles_2000,miller_electrochemical_2008}
Processability and a low cost are essential characteristics when choosing electrode material for supercapacitors. 
That is why one of the most widely used electrode material is graphitic carbon.\cite{frackowiak_carbon_2001} 
To allow for the accumulation of more charge, the specific surface area of graphitic carbon is increased via chemical or thermal activation,\cite{pandolfo_carbon_2006,harmas_influence_2018, tee_electrical_2019} resulting in capacitance as high as 120, 300, and 245~F/g in organic, aqueous, and ionic liquid (IL) electrolytes, respectively.\cite{simon_materials_2008,lust_characteristics_2016} 
In addition to activated carbon,  the potential usage of nanostructured carbon and graphene (Gr) are gaining broad interest.\cite{tallo_nanostructured_2011, zhong_review_2015, yang_new-generation_2016, liu_high_2015}
However, all carbon-based materials have a limitation due to their quantum properties. 
Upon an increase of a specific surface area, the electrode thickness reduces to the size of a space-charge region. 
That, in turn, reduces the number of mobile charge carriers, which decreases the energy storage capacity.\cite{pandolfo_carbon_2006} 
In general, the capacitance of a Gr electrode--electrolyte interface is limited by the capacitance of Gr.\cite{gerischer_density_1987} 
In particular, the electronic structure of Gr directly relates to the characteristic capacitance minimum near the potential of zero charge.\cite{stoller_interfacial_2011}
Still, by choosing a suitable electrolyte, one can enhance the energy storage capacity. 
For example, by widening the electrochemical stability window with the use of an IL electrolyte, the storage capacity of the supercapacitors can be improved. 
That is why thermally stable and moisture resistant room temperature ILs are promising electrolytes for electrochemical application.\cite{fedorov_ionic_2014,macfarlane_energy_2014}

Because macroscopic quantities such as capacitance arise from the microscopical process taking place in the EDL, an understanding of the structure and dynamics of interfaces is crucial for the development of supercapacitors. 
The properties of ILs near surfaces have been actively studied since 2007.\cite{fedorov_ionic_2014,macfarlane_ionic_2007}
Numerous computational studies have focused on the interfacial structure, dynamics, and properties of ILs, using classical molecular dynamics (CMD) \cite{lynden-bell_electrode_2012,vatamanu_molecular_2012,merlet_electric_2014,voroshylova_role_2019,ivanistsev_polymorphic_2014, ivanistsev_restructuring_2015, docampo-alvarez_molecular_2019}, density functional theory (DFT) \cite{valencia_ab_2009, valencia_first-principles_2012, ploger_theoretical_2016,ruzanov_thickness_2018, ruzanov_density_2018}, and DFT-based molecular dynamics (DFT-MD) \cite{ando_electrochemical_2014,paek_influence_2015, yildirim_decomposition_2017,tang_++_2017}. 

For the electrode model, different carbon-based materials were examined; for the electrolyte model, various imidazolium-based ILs were investigated. Specific attention was given to the interface between a Gr sheet and 1-ethyl-3-methylimidazolium tetrafluoroborate (EMImBF$_{4}$) because of the high specific surface area of Gr coupled with the large electrochemical stability window of EMImBF$_{4}$.\cite{nishida_physical_2003} 
Previously, several capacitor models consisting of carbon-based electrode and EMImBF$_{4}$ were studied with CMD by Shim \textit{et al.}\cite{shim_graphene-based_2011}, Merlet \textit{et al.},\cite{merlet_new_2012} and Paek \textit{et al.}\cite{paek_influence_2015}
Shim and co-workers noticed that interfacial capacitance is higher for the positively charged electrode due to the size differences between the ions.
Merlet \textit{et al.} reproduced the results of the earlier atomistic simulation using a coarse-grained model of EMImBF$_{4}$.\cite{merlet_new_2012}
Paek \textit{et al.} conducted, in addition to CMD, a 1.5~ps-long DFT-MD simulation using a model made up of 15 EMImBF$_{4}$ ion pairs and a Gr sheet of 60 atoms. By comparing the potential profiles, Paek \textit{et al.} concluded that the polarisation effects should be included when evaluating the interfacial capacitance.\cite{paek_influence_2015}

In this study, we have examined the Gr--EMImBF$_{4}$ interface with large-scale DFT-MD simulations focusing on its microscopic (electrode charge screening and polarisation) and macroscopic (surface charge density and potential drop) characteristics. 
We show how the analysis of the obtained electronic and geometric structures leads to the estimation of the central quantity -- the potential drop across the EDL.


\section{Computational methods}

\subsection{Classical molecular dynamics simulations}

To run CMD simulations of the Gr--EMImBF$_4$ interface, we used a model with two rigid, parallel Gr sheets with an area of 3.408$\times$3.4433 nm$^2$.
The 10.472~nm space between the sheets was filled with 450 EMImBF$_4$ ion pairs using Packmol\cite{martinez_packmol:_2009}.
All the simulations were run using Gromacs  5.1.4\cite{abraham_gromacs:_2015, pronk_gromacs_2013} software and NaRIBaS scripting framework\cite{roos_nerut_naribasscripting_2018}. 
OPLS-AA force field was used with an effective dielectric constant of 1.6.\cite{sambasivarao_development_2009} 
Each system was pre-equilibrated for 0.1 ns, annealed at 1000, 900, 800~K for 3~ns to produce three replicas, and simulated in the $NVT$ ensemble for 10~ns at 450 K, which was controlled by the $v$-rescale thermostat \cite{bussi_canonical_2007}. 
In between the pre-equilibration and annealing, the electrodes were equally but oppositely charged within 2~ns.
The surface charge densities ($\sigma$) of 0 and $\pm$0.5 e/nm$^2$ were set as point charges of equal magnitude on the surface atoms. 
All other parameters were set the same as in Ref. \citenum{ivanistsev_molecular_2016}.

\subsection{DFT-based molecular dynamics simulations}

\begin{figure}[H]
  \includegraphics{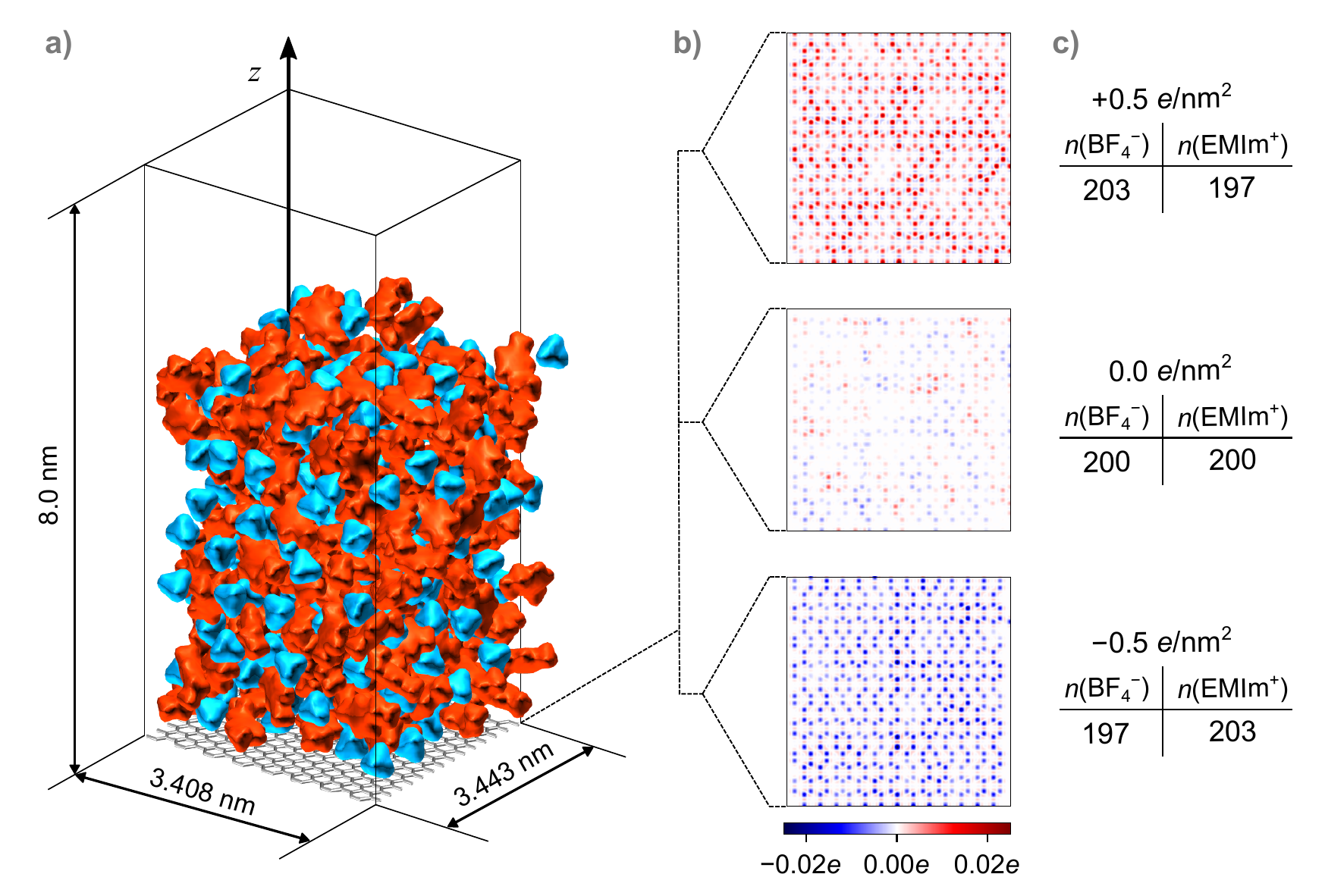}
  \caption{a) Snapshot of the system used within simulations in the current work. The EMIm$^+$ cations, BF$_{4}^{-}$ anions and Gr electrode are colored in red, blue, and gray, respectively. b) The charge distribution of carbon atoms in Gr at different surface charge densities, where each point depicts one carbon atom of Gr. c) The number of ions in the studied systems.}
  \label{fig:cell}
\end{figure}

All model systems used in the DFT-MD simulations were cut out from the last step of the CMD trajectory.
Figure \ref{fig:cell} illustrates the DFT-MD simulation cell, where one side of the cell is the Gr sheet (448 carbon atoms) in contact with IL, which consists of 400 ions. To have a similar $\sigma$ as in the CMD simulations, an excess of EMIm$^+$ cations or BF$_{4}^-$ anions was introduced to cause a natural charge redistribution between IL and Gr. Periodic boundary conditions were applied in $x$,$y$-directions, parallel to the surface. A vacuum layer was added to extend the $z$-axis along the surface normal of Gr from 5 to 8~nm. In this manner, two replicas were created for two systems with the charged electrode and three replicas for the system with the neutral electrode. 
The exact number of ions in each system are given in Figure \ref{fig:cell}. 
The DFT-MD simulations at 353~K, in the $NVT$ ensemble, were performed with CP2k software package versions 2.6 and 5.0 \cite{hutter_cp2k:_2014}, implementing the Gaussian plane wave method \cite{lippert_hybrid_1997}. 
The computations were carried out using the energy cut-off of 600~Ry for the plane wave grid, Perdew--Burke--Ernzerhof (PBE) exchange-correlation functional \cite{perdew_generalized_1996}, optimised double-zeta basis set, with corresponding Goedecker--Teter--Hutter (GTH) pseudopotentials,\cite{goedecker_separable_1996,hartwigsen_relativistic_1998,vandevondele_gaussian_2007} and the D3 dispersion correction by Grimme \textit{et al.} \cite{grimme_consistent_2010}.
It was previously shown in Refs. \citenum{grimme_performance_2012, pensado_effect_2012, bernard_new_2010, izgorodina_importance_2014,karu_performance_2018} that the PBE+D3 combination gives accurate interaction energies for IL ion pairs. 
The duration of the simulations was from 2.3 to 8.1~ps with a time step of 0.5~fs. Within the 0.5--1~ps the bond lengths and angles relaxed from the values given by the CMD force field to the ones consistent with the PBE+D3/GTH/DZVP-MOLOPT theory level.
For further analysis 32--37 random snapshots were chosen after 2~ps of simulation for charge analysis with Chargemol 3.5 software implementing the sixth version of density derived electrostatic and chemical (DDEC6) partitioning approach that shows accurate reproduction of the electrostatic potential for a variety of periodic and non-periodic systems \cite{manz_chemically_2010, manz_introducing_2016,manz_chargemol_2017}. 
The averages over these snapshots were used for the system characterisation. 

\subsection{Density of States calculation}
Unlike in the CMD simulations, DFT calculations explicitly account for the electronic density distribution, which can be used for the calculation of the projected density of states (pDOS). Based on the pDOS of Gr, one can estimate $\sigma$ as
\begin{equation}
    \sigma_\textrm{DOS}=\frac{n(e^{-})}{S}-\int_{-\infty}^{\infty}D(E)P(E)dE,
    \label{eq:8}
\end{equation}
\noindent where $D(E)$ is the projected density of states for Gr at the energy level $E$, $P(E)$ the occupancy of the level, $n(e^{-})$ the number of carbon atoms' valence electrons in case of neutral Gr in vacuum, and $S$ the surface area of Gr in contact with IL.
The additional potential resulting from electronic density redistribution in Gr caused by charging can be estimated from the DOS,\cite{wood_first-principles-inspired_2014}
 
 \begin{equation}
    \Delta\phi_\textrm{DOS}=\frac{E^{0}_\textrm{F}-E^{\sigma}_\textrm{F}}{e},
    \label{eq:6}
\end{equation}

\noindent where $E^{\sigma}_\textrm{F}$ is the Fermi energy of the Gr sheet with a surface charge density $\sigma$, $E^{0}_\textrm{F}$ the energy at the minimum in the DOS shown in Figure~\ref{fig:dip_vs_chg}, and $e$ the elementary charge.

Using the pDOS of Gr, it is also possible to evaluate the Thomas--Fermi screening length of Gr using the equation \cite{castro_neto_electronic_2009}:

\begin{equation}
    l_\textrm{TF}=\frac{\epsilon_{0}\hbar}{4e^{2}}\sqrt{\frac{2E_{F}}{\pi m_{e} n}},
    \label{eq:ThomasFerm2D}
\end{equation}

\noindent where the concentration of mobile charge carriers ($n$) in Gr can be estimated as

\begin{equation}
    n=\int_{-\infty}^{E_\textrm{F}}D(E)[1-f(E-E_\textrm{F},T)] \, dE \ + \int_{E_\textrm{F}}^{\infty}D(E)f(E-E_\textrm{F},T)] \, dE.
    \label{eq:Mobile}
\end{equation}

\noindent In Eq. \ref{eq:Mobile} $f(E-E_\textrm{F},T)$ stands for the Fermi--Dirac distribution function $\left[\exp\left({\frac{E-E_\textrm{F}}{k_\mathrm{B}T}}\right)+1\right]^{-1}$.

\subsection{Overscreening and potential drop}

To quantify the overscreening phenomenon, we used the overscreening factor ($\beta$):
\begin{equation}
    \beta(z)=-\frac{1}{\sigma}\int_{0}^{z}\rho(z')dz',
    \label{eq:1}
\end{equation}
where $\rho_\textrm{ion}(z)$ is the ionic charge density at distance $z$ from the electrode, and $\sigma$ is the surface charge density.
The surface charge is screened when $\beta=1$ and overscreened when $\beta>1$.

To calculate the potential drop, we used the charge density obtained by the means of DDEC6 analysis and Poisson's equation. 
When studying the potential drop in a direction perpendicular to the Gr surface, integration of Poisson's equation yields the expression for electrostatic potential at a point $z$ from the electrode:
\begin{equation}
    \phi(z)=-\frac{1}{\epsilon_0}\int_{0}^{z} (z-z^{'})\rho(z^{'})dz^{'}
    \label{eq:3}
\end{equation}

The potential drop ($\phi_\mathrm{ion}$) between the electrode and the electrolyte was calculated as

\begin{equation}
    \Delta\phi_\textrm{ion}=\phi_\textrm{Gr}-\overline{\phi}{_\textrm{electrolyte}}.
    \label{eq:4}
\end{equation}

\noindent where $\phi_\textrm{Gr}$ is electrostatic potential at the geometric centre of the Gr sheet ($z=0$) and $\overline{\phi}{_\textrm{electrolyte}}$ is an average of electrostatic potential in IL 2.5--4.0~nm from the electrode. 

When estimating the potential drops, instead of mean average, we chose the weighted average approach to reduce the influence of noise on the resulting values. 
First, we calculated the potential drop for each replica by finding the mean average of the potential drops over its snapshots. 
Then, taking the inverse square of the standard deviation of the potential drops of a replica as a weighting factor, we estimated the potential drop of a system with $\sigma$ as the weighted average over the replicas.


\section{Results and discussion}

\subsection{Overscreening phenomena}

One of the main motivations of the presented study was to provide a DFT-level examination of the overscreening phenomena.
While overscreening naturally appears in the CMD, previously it was not clear how it is affected by the polarisation inherent at the DFT level. 
For instance, calculations of Valencia \textit{et al.} indicated a possibility of charge transfer between the contact ions and the surface,\cite{valencia_ab_2009} which in principle should diminish the overscreening.

DDEC6 analysis revealed no charge transfer between the electrode surface and the ions in the first layer -- the charges of anions and cations of $-0.82e$ and $+0.82e$, respectively, were the same for ions in contact with the surface and 2.5--4~nm away from it.
Also, within the length of the obtained trajectories, we did not notice any statistically significant changes in the potential profiles (shown in Figure \ref{fig:allInOne}a) that would indicate the disappearance of the alternate layering.

As an example of density fluctuations, Figure \ref{fig:allInOne}d illustrates how the layering arises due to geometric constraint on ions caused by the neutral Gr surface.\cite{israelachvili_intermolecular_2011} The first layer of the IL, which is considered to start at the van der Waals radius for carbon (0.17 nm from the surface), is 0.40 nm broad and the second layer is 0.42 nm wide. Herewith, near the neutral electrode, the layers contain almost the same number of anions and cations so that the densities of the layers equal the bulk density of 1.24 g/cm$^3$.\cite{fuller_room_1997, nishida_physical_2003}

The densities of the layers become different from the bulk one when the electrode charging causes the overscreening and segregation of anions and cations into distinct layers. Changes in the overscreening parameter $\beta$, shown in Figure \ref{fig:allInOne}b and e, reflect the alternation of ionic layers.
As can be seen in Figures \ref{fig:allInOne}c and f, $\beta$ values are larger than one for both negative and positive electrodes implying dense packing of counter-ions and overscreening.
Due to the smaller size of BF$_{4}^-$, a denser (2.74 ions/nm$^2$) and more ordered structure appears at the positive electrode than at the negative one (1.75 ions/nm$^2$).
Herewith, in all cases the non-uniform ion distribution correlates with the non-uniform surface charge distribution on the Gr electrode as indicated in Figures \ref{fig:cell}b, \ref{fig:allInOne}c and \ref{fig:allInOne}f.

Overall, Figures \ref{fig:allInOne}a--f illustrate that the overscreening phenomena is not an artifact of the CMD simulations as it persists in the more realistic DFT-MD simulations with implicitly included polarisation.

\subsection{Interfacial potential drop}

\begin{figure}[H]
  \includegraphics{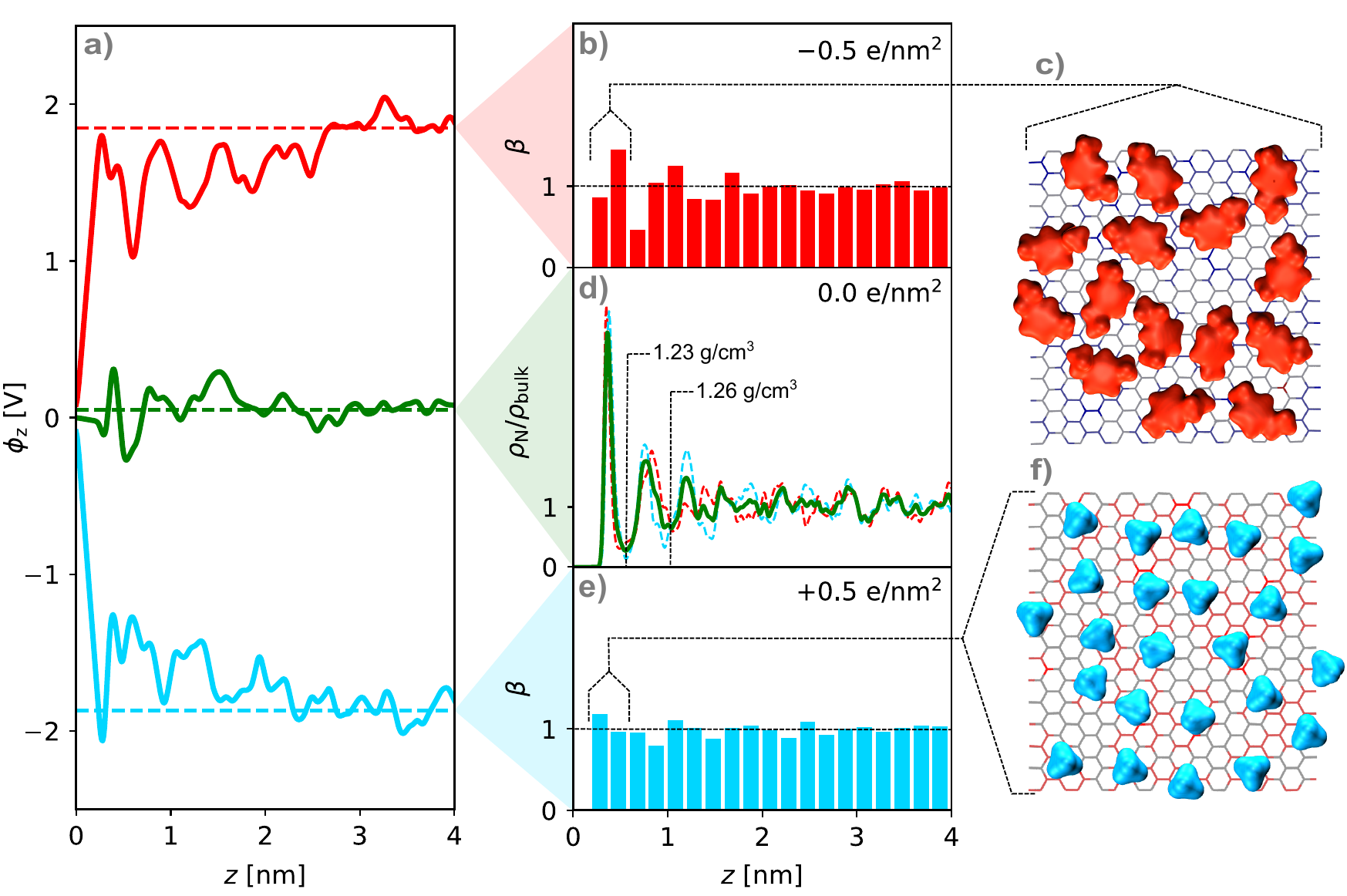}
  \caption{a) The average electrostatic potential ($\phi$) profiles of the systems with $\sigma=$ 0 and $\pm$0.5 e/nm$^2$. b) and e) The average charge screening factor profiles of IL in case of a charged Gr electrode. The bar charts show the averaged values of $\beta$ in 0.15 nm wide bins, to emphasise the dampening of oscillations away from the interface. d) The average number density profile of IL in case of a neutral Gr electrode, where number density is re-scaled by dividing with the number density of the bulk. c) and f) The distribution of ions within 0.4~nm from the charged Gr in one snapshot. The EMIm$^+$ and BF$_{4}^{-}$ ions are colored in red and blue, respectively. }
  \label{fig:allInOne}
\end{figure}

Figure \ref{fig:allInOne}a shows the weighted average potential ($\phi(z)$) profiles in a perpendicular direction to the Gr surface.
As follows from Figure \ref{fig:allInOne}a, the potential drop occurs mostly within 0.3 nm of the Gr surface. 
Then the potential continuously changes until 2.5 nm with pronounced fluctuations. 
At larger distances, the fluctuations dampen and the potential reaches a plateau. 

\begin{figure}[H]
  \includegraphics{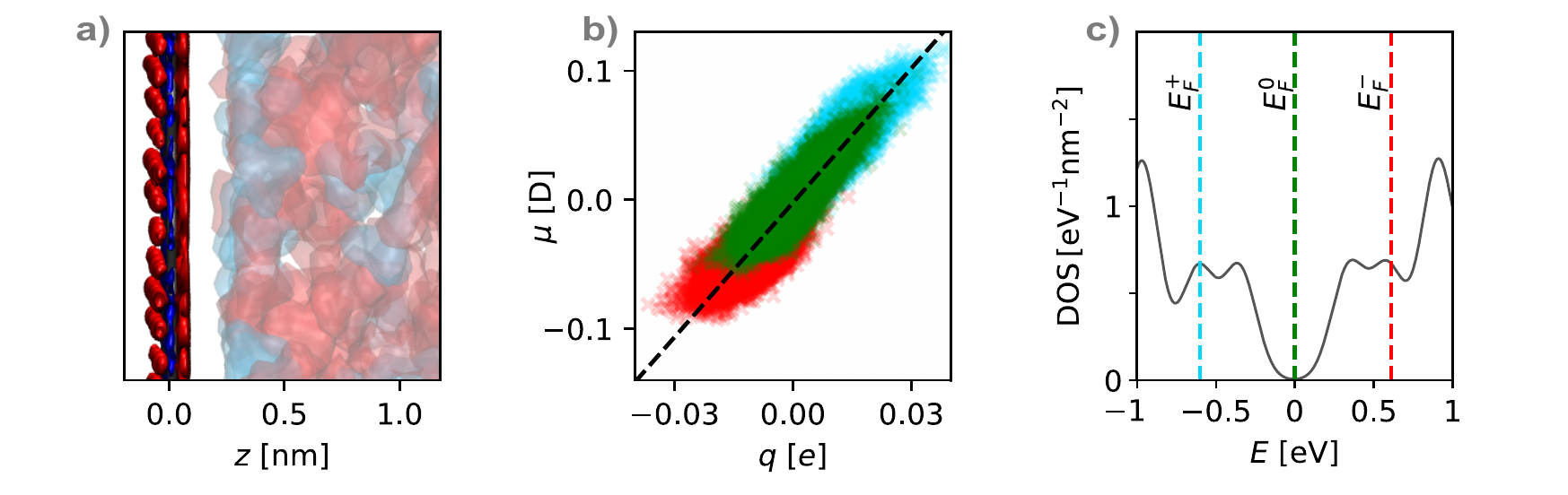}
  \caption{a) Electron density difference at positive Gr surface relative to neutral Gr in a vacuum for a single snapshot. Areas of the electrode, which have lower electron density than neutral Gr are indicated with red color, while higher density is shown with blue color, respectively at $\pm$2.68 e/nm$^3$. IL is represented by transparent red-blue ions. b) The correlation between the partial charges ($q$) and dipole moments of the Gr carbon atoms in a direction perpendicular to surface of the electrode ($\mu$). Red, green, and blue points correspond to negative, neutral, and positive system, respectively. c) Calculated DOS of Gr in contact with IL. The energy is given relative to neutral Gr Fermi level. }
  \label{fig:dip_vs_chg}
\end{figure}

Table \ref{table_pot}  presents values for two corrections to the potential drop that can be evaluated from the DFT data. 
The difference between levels defined in Eq.~\ref{eq:6} and shown in Figure \ref{fig:dip_vs_chg}b describes the additional work done to bring an electron into Gr.
In addition to DOS, the potential change due to electronic density redistribution can also be calculated from the DDEC6 analysis providing dipoles and quadruples in addition to point charges.
The dipoles characterise well the uneven distribution of charge shown in Figure \ref{fig:dip_vs_chg}a. 
Thus, at first approximation, to incorporate the electrode polarisation, the correction term ($\Delta\phi_\textrm{dip}$) can be calculated from the electrode dipole moment in $z$-direction and added to Eq.~\ref{eq:4}:

 \begin{equation}
    \Delta\phi_\textrm{dip}=\frac{p_\textrm{z}}{\epsilon_0 S},
    \label{eq:5}
\end{equation}

\noindent where $p_\textrm{z}$ is the average of dipole moments of the Gr atoms in $z$-direction and $S$ is area of the electrode. 
The comparison between the correction terms in Table \ref{table_pot} reveals a similarity in their values caused by the interrelation between the filling of electron levels and the dipole moment formation. 
Small differences between the correction terms arise due to the constraints of the DDEC6 and DOS analyses.
Furthermore, the potential correction terms can be recalculated into the position of the effective surface charge plane ($l$) along the $z$-axis relative to the center of the electrode: 
 \begin{equation}
    l=\frac{\epsilon_0\Delta\phi}{\sigma}.
    \label{eq:7}
\end{equation}
Theoretically, the surface charge plane should be located one-half of an interplanar spacing. Due low concentration of mobile charge carriers, in Gr the plane is shifted by the Thomas--Fermi screening length towards the surface \cite{lang_theory_1973}. Comparison of the theoretical $a/2-l_\textrm{TF}$ and obtained $l_\textrm{dip/DOS}$ reveals significant difference, yet shows similar $\sigma$-dependence.
\begin{table}
\begin{tabu}{ | r | r | r | r | } 
\cline{1-4}
& \multicolumn{3}{|c|}{$\sigma$} \\
\cline{2-4}
  & $-0.5\phantom{0}$  & $0.0\phantom{0}$  & $+0.5\phantom{0}$ \\ 
\tabucline[2pt]{-}
$\sigma_\textrm{ion}$ & $-0.41$  & $0.005$  & $0.43$ \\ 
\hline
$\phi_\textrm{ion}$  & $-1.87$  & $-0.05$  & $1.85$  \\ 
\hline
$\Delta\phi_\textrm{dip}$ & $-0.65$  & $-0.01$  & $0.60$ \\ 
\hline
$l_\textrm{dip}$ & 0.086  & 0.062  & 0.073 \\
\tabucline[2pt]{-}
$\sigma_\textrm{DOS}$ & $-0.38$  & $0.00$  & $0.31$ \\ 
\hline
$\Delta\phi_\textrm{DOS}$ & $-0.60$  & $0.00$  & $0.60$ \\
\hline
$l_\textrm{DOS}$ & 0.087  & $-0.005$  & 0.105 \\
\hline
$a/2-l_\textrm{TF}$ & 0.16  & 0.05  & 0.15 \\

\hline

\end{tabu}
\caption{Average potential drops $\phi_\textrm{ion}$ and correction terms calculated from dipoles $\Delta\phi_\textrm{dip}$ or Gr density of states $\Delta\phi_\textrm{DOS}$ at different $\sigma$. All $\phi$ are expressed in V, $\sigma$ in elementary charge per square nanometer ($e$/nm$^2$), and all lengths $l_\textrm{TF/dip/DOS}$ in nm.}
\label{table_pot}
\end{table}

To the best of our knowledge, the dipole correction term is defined for the first time in the context of the electrode--IL interface and computer simulations. 
Earlier Kornyshev and Schmickler suggested estimating the effect of electric field on the electrode by evaluating the effective charge density plane from electronic density distribution at the DFT level of theory.\cite{kornyshev_differential_2014} 
A similar approach was used to correct the MD simulations results by Ruzanov \textit{et al.}\cite{ruzanov_thickness_2018} 
Using the DOS, Paek \textit{et al.} accounted for the potential drop within the electrode to express the interfacial capacitance as $1/C_{\textrm{total}} = 1/C_{\textrm{Gr}} + 1/C_{\textrm{IL}}$.\cite{paek_computational_2013}
Dufils \textit{et al.} also used the DOS to evaluate Thomas--Fermi screening length, which they used to tune the metallicity of electrodes in MD simulations \cite{dufils_semiclassical_2019}.
The suggested dipole correction provides an alternative way to evaluate the potential drop within the electrode.
Similarly to the mentioned works \citenum{kornyshev_differential_2014,ruzanov_thickness_2018,paek_computational_2013,dufils_semiclassical_2019}, the dipole correction can be expressed as an average quantity for a surface as a whole.
In that case, a separate calculation is required to obtain a value of dipole correction for each field, potential, or surface charge value.
Alternatively, a correlation between the dipole moment and charge on individual surface atoms (shown in Figure \ref{fig:dip_vs_chg}c) can be used.
Atomic charges can be already self-consistently calculated using the constant potential method, for example developed or used in Refs. \cite{li_computer_2018, merlet_simulating_2013, haskins_evaluation_2016, wang_evaluation_2014, vatamanu_application_2017, breitsprecher_electrode_2015, coles_simulation_2019, scalfi_charge_2020}.
Applying the dipole correction over the atomic charges opens a door for the low-cost augmentation of the CMD simulations to account for the electrode polarisation.

\section{Conclusions}
In this work, we used DFT-based MD to explore the Gr--1-ethyl-3-methyl-imidazolium tetrafluoroborate interface to examine the charge distribution and polarisation on the Gr surface accounting for the dynamics of ions near the surface. The results indicate that overscreening is preserved upon the inclusion of quantum effects. No indication of the charge transfer between the IL and the electrode was seen and the layered structure in the IL near the Gr surface remained stable during up-to 8~ps long DFT-based MD simulation.

The analysis showed that the atomic charges on the electrode are distributed non-uniformly due to electronic lateral polarisation of the electrode. Polarisation and redistribution of Gr electron density in the direction perpendicular to the surface induces a surface dipole that significantly affects the interfacial potential drop.

To reconcile the differences between classical and DFT-based MD simulations, we introduced a correction term, which quantifies the redistribution of the Gr electron density using atomic dipoles that correlate with the atomic charges. That correction term accounts for the effect of polarisation on the interfacial potential drop and can be used in classical molecular dynamics simulations.

\section{Acknowledgements}
This research was supported by the Estonian Research Council (grants IUT20-13, PSG249, and PUT1107), the EU through the European Regional Development Fund (TK141 ``Advanced materials and high-technology devices for energy recuperation systems''), and the Estonian--French cooperation program Parrot, funded by the Estonian Research Council and Campus France. For providing us with the computational resources, we would like to acknowledge the Partnership for Advanced Computing in Europe (PRACE), the Distributed European Computing Initiative (DECI), the HPC Center of the University of Tartu, and Skoltech Pardus HPC cluster.


\bibliography{prace-jpcl}


\end{document}